\documentclass[twocolumn,times]{aastex63}

\usepackage{amsmath}

\newcommand{\kms}{km\,s$^{-1}$}

\newcommand{\bz}{$\langle B_{\rm z} \rangle$}
\newcommand{\bs}{$\langle B \rangle$}
\newcommand{\au}{AU\,Mic}

\newcommand{\fifps}[2]{\centering\resizebox{#1}{!}{\includegraphics{#2}}}

\newcommand{\mybf}[1]{{#1}}

\shorttitle{Magnetic Field of AU\,Mic}

\shortauthors{Kochukhov \& Reiners}

\begin{document}

\title{Magnetic Field of the Active Planet-hosting M Dwarf AU\,Mic}

\correspondingauthor{Oleg Kochukhov}
\email{oleg.kochukhov@physics.uu.se}

\author[0000-0003-3061-4591]{Oleg~Kochukhov}
\affil{Department of Physics and Astronomy, Uppsala University, Box 516, 75120 Uppsala, Sweden}

\author[0000-0003-1242-5922]{Ansgar~Reiners}
\affil{Institut f\"ur Astrophysik, Georg-August-Universit\"at, Friedrich-Hund-Platz 1, 37077 G\"ottingen, German}

\begin{abstract}
AU~Mic is a young, very active M dwarf star with a debris disk and at least one transiting Neptune-size planet. Here we present detailed analysis of the magnetic field of AU~Mic based on previously unpublished high-resolution optical and near-infrared spectropolarimetric observations. We report a systematic detection of circular and linear polarization signatures in the stellar photospheric lines. Tentative Zeeman Doppler imaging modeling of the former data suggests a non-axisymmetric global field with a surface-averaged strength of about 90~G. At the same time, linear polarization observations indicate the presence of a much stronger $\approx$\,2~kG axisymmetric dipolar field, which contributes no circular polarization signal due to the equator-on orientation of AU~Mic. A separate Zeeman broadening and intensification analysis allowed us to determine a mean field modulus of 2.3 and 2.1~kG from the Y- and K-band atomic lines respectively. These magnetic field measurements are essential for understanding environmental conditions within the AU~Mic planetary system.
\end{abstract}



\section{Introduction}
\label{sec:intro}

Magnetism plays a prominent role in the physics of low-mass stars and their planetary systems. M dwarfs are known to exhibit copious evidence of magnetically driven surface activity in the form of flares, photometric rotational modulation, chromospheric and coronal emission \citep[e.g.][]{hawley:2014,newton:2016,newton:2017,astudillo-defru:2017,wright:2018}. Direct magnetic field measurements demonstrate that these objects possess some of the strongest fields found in cool active stars \citep{morin:2008,kochukhov:2017c,shulyak:2017}. M-dwarf surface magnetic field geometries have complex, multi-component nature \citep{reiners:2009,kochukhov:2019a,afram:2019} with a moderately strong, often axisymmetric, global field component measured with high-resolution circular polarimetry \citep{donati:2006,donati:2008,morin:2008,morin:2010,lavail:2018} superimposed onto a much stronger tangled small-scale field that gives rise to Zeeman broadening and splitting in the intensity spectra \citep{saar:1985,johns-krull:1996,reiners:2006a,kochukhov:2009b,shulyak:2017,shulyak:2019}. The relationship between these two magnetic components is currently poorly characterized as a function of stellar rotation and fundamental parameters and is not reproduced by theoretical models of dynamo operating in fully convective stars \citep{dobler:2006,browning:2008,yadav:2015}.

M-dwarf magnetism has dramatic consequences for the close-in terrestrial planets orbiting these stars. Global magnetic fields of low-mass stars modulate their X-ray activity \citep{cook:2014,lang:2014}, govern propagation of the coronal mass ejections \citep{lynch:2019}, determine the stellar wind pressure exerted on planets \citep{vidotto:2011,vidotto:2013,garraffo:2016}, influence the chemistry and energy balance of the planetary atmospheres \citep{cohen:2014}, and even affect the rocky planet interiors \citep{kislyakova:2017,kislyakova:2018}. Consequently, characterization of stellar magnetic field has become a staple ingredient of any study of exoplanets orbiting low-mass stars. Consideration of the magnetic activity is particularly relevant for young systems in which the host star has not yet shed its angular momentum and thus generates more intense magnetic fields than found in older slower rotating M dwarfs.

\au\ (M1Ve, GJ~803, HD~197481) is one of the nearest and best studied young M dwarfs. It has an age of 22--24~Myr estimated from the membership in the $\beta$~Pic moving group \citep{mamajek:2014,bell:2015}. Due to its proximity \citep[$d=9.72\pm0.04$~pc,][]{gaia-collaboration:2018}, the circumstellar environment of \au\ can be spatially resolved with various high-contrast imaging methods. These observations reveal an edge-on debris disk \citep{kalas:2004,strubbe:2006,grady:2020} with a complex and dynamic clump structure \citep{boccaletti:2015,boccaletti:2018}. The transiting planet, AU~Mic~b, was reported by \citet{plavchan:2020}. This Neptune-size planet has an orbital period of 8.46~d and a semi-major axis of about 19 stellar radii. Thus, it is located sufficiently close to the host star to be significantly affected by its magnetic field. 

Motivated by these discoveries, we present in this paper a detailed analysis of the surface magnetic field of \au\ using high-resolution polarimetry and spectroscopy. Our aim is to characterize both the global and small-scale stellar magnetic field components in order to inform future studies of \au\ and its young planetary system.

\section{Observational data}
\label{sec:obs}

\subsection{Optical spectropolarimetry}

\subsubsection{ESPaDOnS observations}
\label{sec:esp}

Several high-resolution circular and linear polarization observations of \au\ were obtained with the ESPaDOnS instrument \citep{manset:2003} at the Canada-France-Hawaii Telescope (CFHT) on Maunakea, Hawaii. These spectra were retrieved from the PolarBase archive \citep{petit:2014}. Four very high-quality Stokes $V$ spectra were collected for this star on July 15--16, 2005 using 3600~s exposures for each observation. Three more Stokes $V$ spectra were secured on August 2--3, 2006 with 4800--7200~s exposure times. Two Stokes $Q$ and two Stokes $U$ measurements were also made during the latter observing run using 2400~s exposure times. Also, a single linear polarization measurement, not accompanied by a Stokes $V$ spectrum, was collected on May 9, 2006. In this case, 2160~s exposure times were employed for Stokes $Q$ and $U$ observations.

ESPaDOnS spectra cover the 3691--10482~\AA\ wavelength interval and have a resolution of $R\approx65\,000$. The data provided by PolarBase are fully reduced, including continuum normalisation. The ESPaDOnS observations of \au\ are characterised by a signal-to-noise ratio ($S/N$) of 500--600 per 1.8~\kms\ spectral pixel in the 6000--6500~\AA\ wavelength region. The log of the Stokes $V$ ESPaDOnS observations of \au, including UT observing dates, the heliocentric Julian dates (HJD) of mid-exposure, and individual $S/N$ values, is provided in the upper part of Table~\ref{tab:obs}.

\subsubsection{HARPSpol observations}

Five circular polarization observations of \au\ were made with the HARPSpol polarimeter \citep{piskunov:2011,snik:2011} attached to the HARPS spectrometer \citep{mayor:2003} fed by the ESO 3.6m telescope at La Silla, Chile. One observation was obtained on May 3, 2010. Four more Stokes $V$ spectra were taken on different nights from August 8 to August 13, 2010. In all cases, an exposure time of 3600~s was used, yielding a median $S/N$ of 44--188 per 0.8~\kms\ spectral pixel in the 6000--6500~\AA\ wavelength region.

HARPSpol spectra cover the 3780--6913~\AA\ wavelength region with a small gap around 5300~\AA. The resolution is $R$\,$\approx$\,110\,000. The \au\ observations analysed here were processed with the REDUCE pipeline \citep{piskunov:2002} as described in detail elsewhere \citep{makaganiuk:2011,makaganiuk:2011a}. Similar to the ESPaDOnS observations described above, each HARPSpol spectropolarimetric measurement comprised four sub-exposures obtained at different orientations of the quarter-wave retarder waveplate relative to the beamsplitter. The Stokes $V$ parameter spectrum and the diagnostic null spectrum were derived following the ``ratio method'' demodulation scheme \citep{donati:1997,bagnulo:2009}, which allows one to efficiently suppress spurious and instrumental polarization.

Information on individual HARPSpol observations of \au\ is given in the bottom part of Table~\ref{tab:obs}.

\begin{table}
\centering
\caption{Journal of circular polarization observations of \au.}
\label{tab:obs}
\begin{tabular}{llrlr}
\hline
\hline
UT date &   HJD   & $S/N$ & $\sigma_{\rm LSD}$ &  \bz\ (G) \\       
\hline
\multicolumn{5}{c}{\it EPSaDOnS observations} \\
2005-07-16 & 2453567.9523 & 654 & 3.86e-5 &  $21.2\pm1.4$ \\          
2005-07-16 & 2453568.0131 & 551 & 4.46e-5 &  $12.8\pm1.7$ \\          
2005-07-17 & 2453568.9141 & 590 & 4.05e-5 & $-30.2\pm1.5$ \\          
2005-07-17 & 2453569.0061 & 605 & 3.90e-5 & $-27.5\pm1.5$ \\          
2006-08-02 & 2453949.9043 & 466 & 5.38e-5 &  $46.3\pm2.0$ \\          
2006-08-02 & 2453950.0377 & 527 & 5.37e-5 &  $69.1\pm2.0$ \\          
2006-08-03 & 2453950.9282 & 480 & 4.75e-5 &  $34.6\pm1.7$ \\          
\hline
\multicolumn{5}{c}{\it HARPSpol observations} \\
2010-05-03 & 2455319.8986 & 188 & 0.83e-4 & $  5.4\pm2.2$ \\    
2010-08-08 & 2455416.8726 &  91 & 1.74e-4 & $ 53.0\pm4.6$ \\    
2010-08-09 & 2455417.8745 &  44 & 3.63e-4 & $-34.4\pm9.5$ \\    
2010-08-10 & 2455418.8394 &  97 & 1.61e-4 & $ -2.9\pm4.2$ \\    
2010-08-13 & 2455421.8552 & 130 & 1.17e-4 & $ 50.2\pm3.1$ \\    
\hline
\end{tabular}
\end{table}

\subsection{Near-infrared spectroscopy}

\subsubsection{Y-band observations}

High-resolution spectra of \au\ in the Y band ($\lambda$ 9640--9790~\AA) were employed in this study to measure the mean magnetic field modulus using magnetically sensitive atomic lines. In principle, any ESPaDOnS observation described in Sect.~\ref{sec:esp} is suitable for this analysis. However, considering that telluric lines are strong in the region of interest, we made use of the two spectra taken on the night of May 9, 2006 (HJD 2453896.1032 and 2453896.1307) because the telluric absorption was significantly less intense in these observations and was shifted away from the key stellar diagnostic lines. These observations were carried out in the linear polarization mode using 2160~s exposure times. 

The data extracted from PolarBase were post-processed by removing telluric lines. For this purpose, we made use of the slightly adjusted theoretical telluric spectrum calculated by the TAPAS service \citep{bertaux:2014} for the atmospheric conditions corresponding to the \au\ observations at CFHT. The final co-added Stokes $I$ spectrum of \au\ has $S/N\approx500$ per 1.8~\kms\ pixel in the  9640--9790~\AA\ region.

\subsubsection{K-band observations}

A measurement of the mean magnetic field of \au\ was also performed using high-resolution K-band spectra collected with the CRIRES instrument \citep{kaeufl:2004} mounted at the 8m ESO VLT. These observations, carried out on May 8, 2012 (HJD 2456055.9098), cover the 22048--22487~\AA\ wavelength region in four short segments recorded on individual detectors. Here we use only the 22202--22318~\AA\ interval corresponding to the second CRIRES detector. This region contains several magnetically sensitive Ti~{\sc i} lines commonly used for the Zeeman broadening analysis of low-mass stars. 

Eight 30~s sub-exposures acquired at two nodding positions along the slit resulted in $S/N\approx700$ per 1.5~\kms\ pixel in the combined extracted spectrum. The nominal resolving power of these observations, obtained with a slit width of 0.2\arcsec\ and with an adaptive optics system active, is $R\approx10^5$. The CRIRES spectra were reduced with the help of the standard ESO pipeline, as described by e.g. \citet{shulyak:2014}. Similar to the processing of the Y-band data, a custom theoretical telluric spectrum was calculated with TAPAS for the specific atmospheric conditions of the \au\ observations at VLT. These calculations were first employed to improve the wavelength calibration of the CRIRES spectrum and then to remove telluric spectral features.

\begin{figure*}[!th]
\fifps{0.63\hsize}{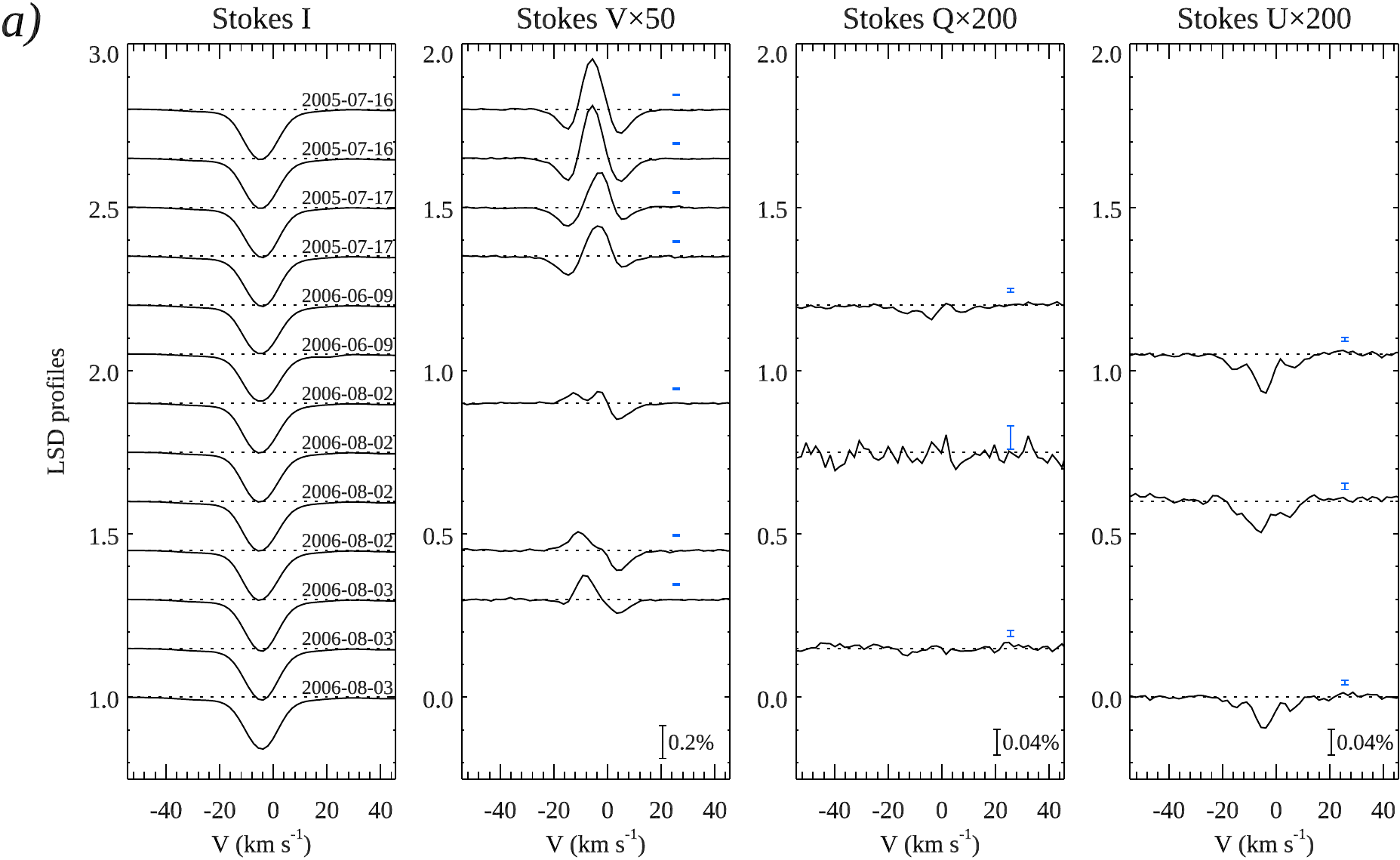}\hspace*{5mm}
\fifps{0.33\hsize}{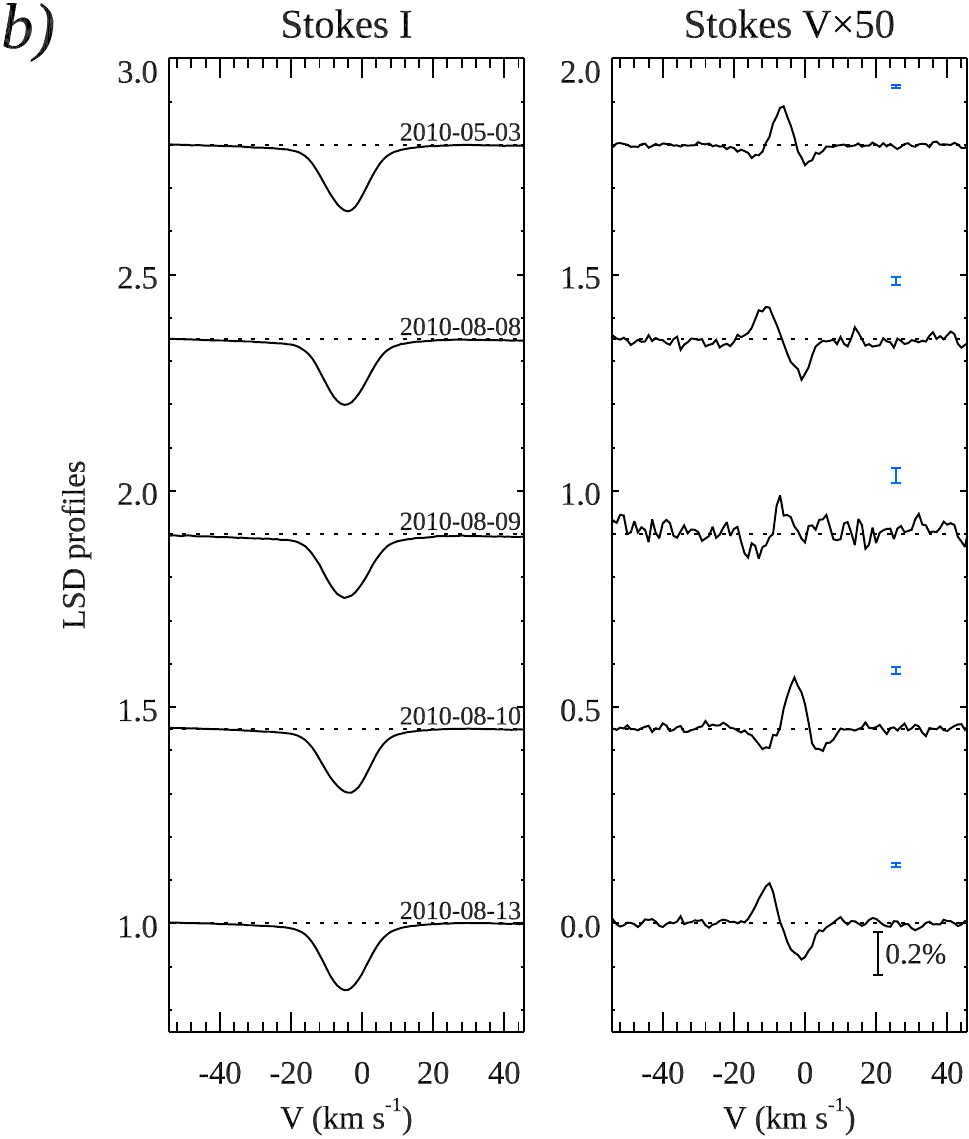}
\caption{LSD Stokes parameter profiles of \au\ derived from ESPaDOnS (a) and HARPSpol (b) spectropolarimetric observations. Profiles corresponding to different observations are shifted vertically. The polarization profiles are scaled up relative to Stokes $I$ by the factors indicated above each panel.}
\label{fig:lsd}
\end{figure*}

\section{Magnetic field analysis}
\label{sec:mag}

\subsection{Global magnetic field}
\label{sec:zdi}

\subsubsection{Least-squares deconvolution}
\label{sec:lsd}

We applied the least-squares deconvolution \citep[LSD,][]{donati:1997,kochukhov:2010a} multi-line technique to derive high-quality mean circular and linear polarization profiles from the ESPaDOnS and HARPSpol observations of \au. \mybf{To this end, we made use of the iLSD code described by \citet{kochukhov:2010a}. The line mask required by this software} was constructed from an atomic line list retrieved from the VALD data base \citep{ryabchikova:2015} with the atmospheric parameters $T_{\rm eff}=3700$~K and $\log g=4.5$. We retained only lines deeper than 20\% of the continuum and excluded regions affected by telluric absorption. Stellar lines significantly deviating from the average shape (e.g. emission lines, strong metal lines with broad wings) were also excluded. The final mask contained 5331 lines for ESPaDOnS (mean wavelength $\lambda_0=5800$~\AA, mean effective Land\'e factor $z_0=1.2$) and 5108 lines for HARPSpol spectra ($\lambda_0=5300$~\AA, $z_0=1.2$). The resulting typical error bar (polarimetric sensitivity) was $5\times 10^{-5}$ per 1.8~\kms\ velocity bin for the ESPaDOnS LSD profiles and $10^{-4}$ per 1.0~\kms\ bin for the LSD profiles derived from HARPSpol observations (see column 4 in Table~\ref{tab:obs}).

With this multi-line analysis we have succeeded in detecting circular polarization signatures with an amplitude of 0.1--0.3\% relative to the Stokes $I$ continuum in all 12 Stokes $V$ observations of \au. Moreover, linear polarization signals with an amplitude of 0.05\% were systematically detected in the ESPaDOnS Stokes $U$ observations. These signatures appear to have approximately constant shape and amplitude. One marginal polarization signature, with an amplitude less than 0.02\%, was also detected in the Stokes $Q$ spectrum from May 6, 2009. Two other Stokes $Q$ observations showed no signatures associated with the stellar photospheric spectrum.  

All LSD Stokes parameter profiles of \au\ are illustrated in Fig.~\ref{fig:lsd}. There is a significant variability of both the shape and amplitude of the Stokes $V$ profiles. On the other hand, the Stokes $U$ profile is largely stationary. The seven ESPaDOnS Stokes $V$ observations come in three batches separated by 2--11 months. Each group covers at most two consecutive nights at a time. It is difficult to tell from these data if the observed variation of the Stokes $V$ profiles is due to rotational modulation or caused by a long-term magnetic activity cycle. The HARPSpol data set is more informative in this respect. The four Stokes $V$ profiles corresponding to the observations obtained on August 8--13, 2010 (Fig.~\ref{fig:lsd}b) exhibit significant variability and include all basic profile shape types seen in the ESPaDOnS data set. This indicates that the observed Stokes $V$ profile changes are likely explained by the rotational modulation and that the global field topology of \au\ is dominated by a non-axisymmetric component.

Interestingly, some of the strongest circular polarization profiles in both data sets (July 16--17, 2005 for ESPaDOnS, May 3 and August 10, 2010 for HARPSpol) correspond to the symmetric, W-shaped Stokes $V$ profiles. This so-called crossover signature \citep{mathys:1995a} is explained by the presence of a pair of regions with opposite magnetic polarity at the approaching and receding stellar hemispheres, which again suggests a non-axisymmetric large-scale field geometry. The other Stokes $V$ profiles exhibit the classical anti-symmetric S-shape corresponding to a positive line of sight magnetic field component. One HARPSpol Stokes $V$ spectrum (August 9, 2010) shows a weak negative circular polarization signature. The HARPSpol observations obtained five nights apart (August 8 and August 13, 2010) yield very similar Stokes $V$ LSD profiles, which is consistent with recent estimates of the stellar rotation period $P_{\rm rot}$\,$\approx$\,4.85~d \citep{ibanez-bustos:2019,plavchan:2020}.

We quantified the strength of the disk-averaged line of sight (longitudinal) magnetic field \bz\ by computing the first moment of the Stokes $V$ profile and normalising by the equivalent width of the Stokes $I$ profile \citep{kochukhov:2010a}. The measurement window of $\pm20$~\kms\ from the mean radial velocity of \au\ was adopted. This analysis yielded 12 \bz\ measurements reported in the last column of Table~\ref{tab:obs}. We found longitudinal field in the range from $-34$ to $69$~G with typical error bars of 1.5--2~G for ESPaDOnS and 2--9~G for HARPSpol data. Note that, although \bz\ is formally compatible with zero for the HARPSpol observations on May 3 and August 10, 2010, the crossover LSD polarization signatures are still detected unambiguously on both nights.

The diagnostic null spectra were processed with the same LSD approach as was applied to the Stokes $V$ and Stokes $QU$ observations. No polarization signatures were detected in any of the null LSD profiles. This confirms that the line polarization signatures found for \au\ are likely to be of stellar origin and are not significantly affected by instrumental artefacts. The basic difference in shape of the symmetric Stokes $U$ signal from the anti-symmetric Stokes $V$ profile corresponding to the data obtained on the same nights indicates that the Stokes $U$ signature could not have been produced by a cross-talk from Stokes $V$. 

The three-lobe morphology of the Stokes $U$ signature, clearly discernible in two out of three linear polarization observations, is qualitatively consistent with the profile shape expected for the Zeeman effect for a line with triplet magnetic splitting. Comparison of the Stokes $U$ LSD profiles calculated using sub-sets of lines with different magnetic sensitivity shows that the linear polarization amplitude correlates with the effective Land\'e factor. This reinforces the conclusion that the observed Stokes $U$ signature originates from the Zeeman effect and is linked to the stellar magnetic field.

\subsubsection{Zeeman Doppler imaging}

Considering the scatter of recent determinations of the rotation period of \au\ \citep{ibanez-bustos:2019,plavchan:2020}, 
we estimate its precision to be approximately 0.01--0.02~d.
This uncertainty, together with the possibility of intrinsic changes of the surface magnetic field geometry, precludes modeling of the entire collection of \au\ spectropolarimetric observations within a single framework. The five HARPSpol spectra taken in 2010 over about 100 days are better suited for that purpose. Therefore, in this section we interpret these data with the Zeeman Doppler imaging \citep[ZDI,][]{kochukhov:2016} technique. Due to spareness of the rotational phase coverage, the HARPSpol observations are hardly sufficient for deriving a robust global magnetic field model. Nevertheless, we attempted to reproduce these four Stokes $V$ profiles with the ZDI code InversLSD \citep{kochukhov:2014} in order to get a rough idea about the structure and strength of the global magnetic field of \au.

\begin{figure}[!th]
\fifps{\hsize}{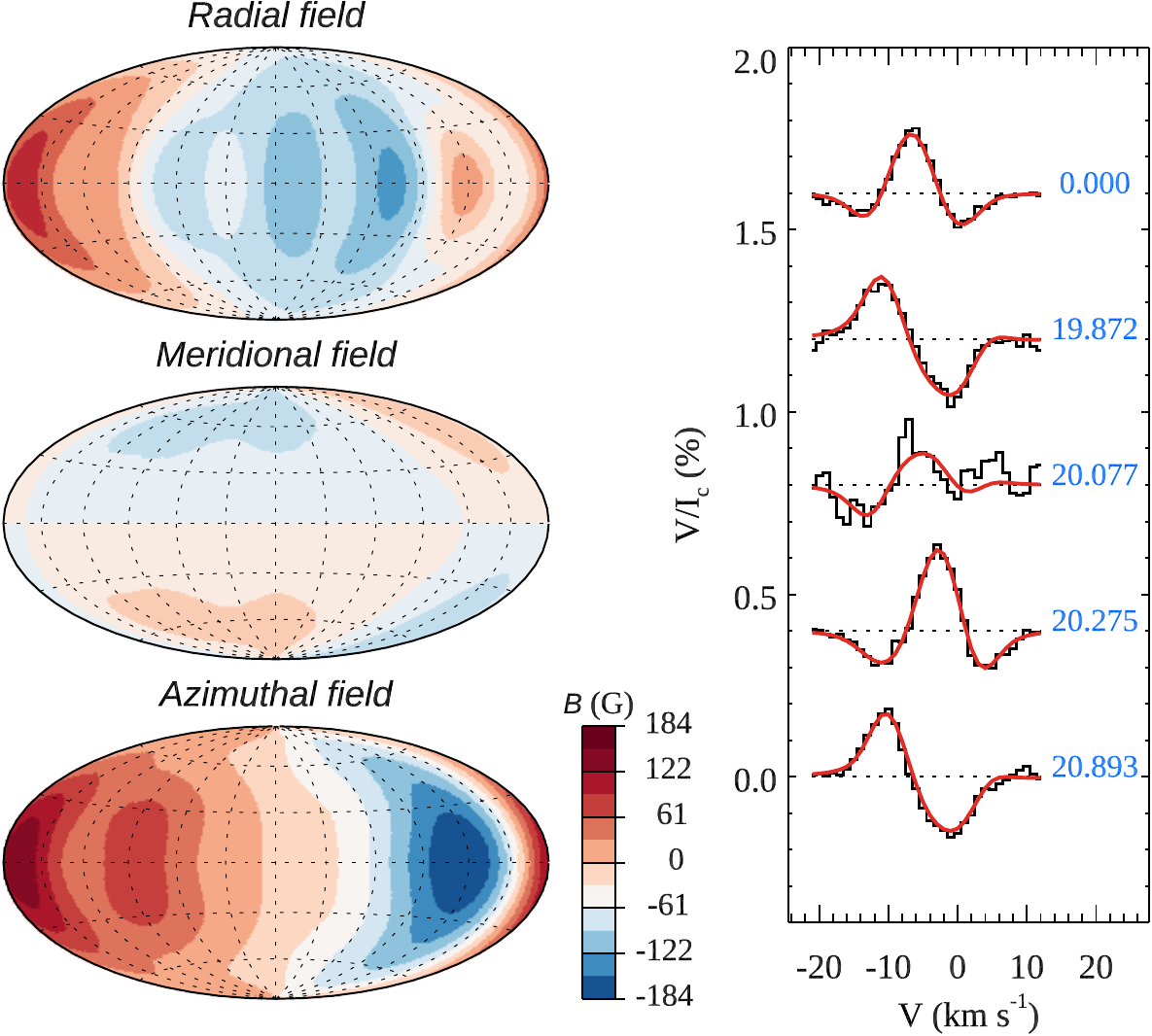}
\caption{Results of ZDI fit to the four HARPSpol Stokes $V$ observations of \au\ obtained in August 2010. The radial, meridional, and azimuthal magnetic field maps are displayed on the left using the Hammer-Aitoff projection. Comparison of the observed (histogram) and model (solid lines) circular polarization spectra is presented on the right. The profiles corresponding to different observing dates are offset vertically. Rotational phases are calculated with the ephemeris $HJD = 2455319.8986 + 4.88 \times E$.}
\label{fig:zdi}
\end{figure}

The ZDI analysis was carried for \au\ similar to previous applications of InversLSD to G, K, and M stars \citep{kochukhov:2015a,hackman:2016,rosen:2016,lavail:2018}. The local Stokes parameter profiles were approximated with the Unno-Rachkovsky analytical solution of the polarized radiative transfer equation \citep[e.g.][]{polarization:2004}. The line strength and broadening parameters were adjusted to match the Stokes $I$ LSD profile. The projected rotational velocity $v_{\rm e}\sin i$ of 9.17~\kms\ was used according to the results of Sect.~\ref{sec:intens} and the inclination angle of the stellar rotational axis was set to 90\degr, consistent with the observation of spin-orbit alignment \citep{martioli:2020} and the edge-on orientation of the debris disk \citep{watson:2011,greaves:2014}. 

The first HARPSpol observation in the series was taken about 20 stellar rotations prior to the remaining four Stokes $V$ observations. In this case, a period error of 0.01--0.02~d incurs about 0.2--0.4 uncertainty of the rotational phase. This necessitates adjusting rotational period to achieve a better match to the spectropolarimetric observations, as commonly practiced by ZDI studies of M dwarfs \citep{donati:2008,morin:2008}. We found that $P_{\rm rot}=4.88$~d minimizes the chi-square of the fit to Stokes $V$ profiles. This value of the rotational period, reasonably consistent with other determinations, was adopted for the final magnetic inversion.

Similar to other modern applications of ZDI, we represented the stellar magnetic field topology with a spherical harmonic expansion \citep{kochukhov:2014} that included both poloidal and toroidal terms. The expansion was truncated at the angular degree $\ell_{\rm max}=5$. However, the equator-on orientation of \au\ leads to several degeneracies in applications of surface mapping techniques. Specifically, for the spherical harmonic global field description adopted in our study, contribution of any axisymmetric harmonic mode which is anti-symmetric with respect to the stellar equator (i.e. any odd-$\ell$, $m=0$ spherical harmonic) is cancelled out in circular polarization observables independently of the quality of observations and rotational phase coverage. This means that contributions of these field components, including a dipole aligned with the stellar rotational axis, cannot be determined from Stokes $V$ observations.

Figure~\ref{fig:zdi} presents results of our tentative ZDI modeling. The magnetic field maps reveal a distinctly non-axisymmetric field structure, with the local field strength reaching 184~G. The average global field strength is 88~G. The harmonic decomposition of the field indicates that the poloidal component dominates over the toroidal one (91.2\% of the magnetic field energy is in the poloidal field) and that the large-scale field is almost entirely non-axisymmetric (96.0\% of the magnetic energy is in $|m| \ge \ell/2$ modes). The dipolar modes contribute 70.2\% of the magnetic energy. These quantitative characteristics of the global magnetic field of \au\ are summarised in the upper part of Table~\ref{tab:mag}.

\subsubsection{Constraints on axisymmetric field component}

A non-axisymmetric global magnetic field structure of \au\ inferred from the qualitative analysis of Stokes $V$ profiles and suggested by the preliminary ZDI map may be incomplete 
\mybf{due to two separate issues. First, the phase coverage of the HARPSpol spectropolarimetric data is sparse and some surface structures might have been missed owing to a large phase gap between observations corresponding to the rotational phases 0.27 and 0.87. Second, our magnetic inversion suffers from a more fundamental limitation.}
For the stellar inclination $i\approx90\degr$, any field component with opposite field polarities above and below the stellar equator, for example a dipole aligned with the rotational axis, will not produce a measurable circular polarization. However, this degeneracy can be alleviated with the help of linear polarization observations. 
A comparison of linear and circular polarization is meaningful only if the measurements are done almost simultaneously.
Fortunately, two complete Stokes $VQU$ sequences were obtained for \au\ on August 2--3, 2006. Based on the LSD profiles corresponding to these full Stokes vector spectra we conclude that the ratio of the maximum absolute Stokes $V$ amplitude $|V|_{\rm max}$ to the maximum total linear polarization $(P_{\rm L})_{\rm max}\equiv (\sqrt{Q^2+U^2})_{\rm max}$ is 2.3--3.1. One can use this measurement to assess the strength of axisymmetric dipolar field that cannot be diagnosed with Stokes $V$ alone.

\begin{figure}[!th]
\fifps{0.9\hsize}{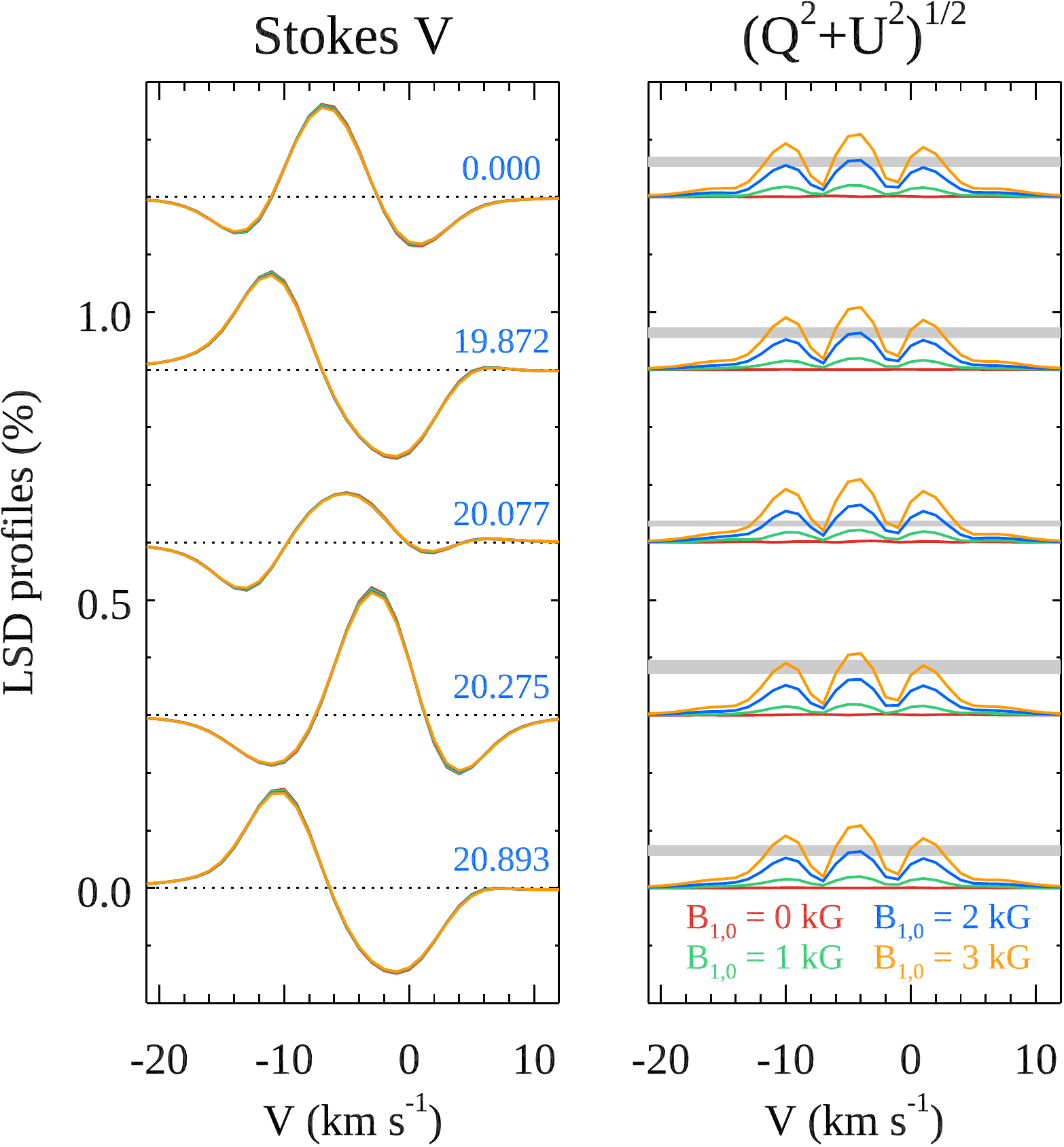}
\caption{Theoretical Stokes $V$ (left) and total linear polarization $P_{\rm L}=\sqrt{Q^2+U^2}$ (right) LSD profiles for different strengths of the axisymmetric dipolar field component. The profiles corresponding to different rotation phases are offset vertically. The horizontal grey bar in the right panel shows the observed linear polarization amplitude estimated from the EPSaDOnS Stokes $QU$ observations.
}
\label{fig:qu}
\end{figure}

Fig.~\ref{fig:qu} shows a series of forward four Stokes parameter calculations with InversLSD for the same four rotational phases of HARPSpol spectra as we modeled with ZDI in the previous section. This figure compares synthetic $V$ and $P_{\rm L}$ profiles computed for the original non-axisymmetric global field map (Fig.~\ref{fig:zdi}) with the calculations for the maps in which the $\ell=1$, $m=0$ harmonic component was set to $B_{1,0}=1$, 2, and 3 kG. In accordance with the discussion above, this modification has no influence on Stokes $V$, but linear polarization increases with $B_{1,0}$. The observed $|V|_{\rm max}/(P_{\rm L})_{\rm max}$ ratio is reproduced with a dipolar field strength of 1.3--2.7~kG. Considering only the HARPSpol spectra for the rotational phases \mybf{0.872 and 0.893} (August 8 and 13, 2010), which exhibit very similar Stokes $V$ signatures in comparison to August 2--3, 2006 ESPaDOnS observations, we can reduce this range to $B_{1,0}$\,$=$\,1.8--2.2~kG. Thus, we find it plausible that \au\ possesses a fairly strong axisymmetric dipolar field in addition to the non-axisymmetric large-scale field of the type illustrated in Fig.~\ref{fig:zdi}.

\subsection{Mean field modulus}
\label{sec:intens}

\subsubsection{Y-band analysis}
\label{sec:yband}

Previous studies \citep{kochukhov:2017c,kochukhov:2019a,shulyak:2017,shulyak:2019} have established the group of ten Ti~{\sc i} lines at $\lambda$ 9647.37--9787.69~\AA\ as a very useful indicator of magnetic fields in low-mass main sequence stars. These neutral titanium lines belong to the same multiplet, which eliminates uncertainty of their modeling stemming from oscillator strength errors and ensures that they form under similar thermodynamic conditions in stellar atmospheres. These lines exhibit a widely different response to a magnetic field. Several of them (e.g. $\lambda$ 9647.37, 9688.87, 9770.30, the blend of 9783.31 and 9783.59, and 9787.69~\AA) are noticeably strengthened and broadened by the Zeeman effect whereas other lines are affected to a lesser extent. This multiplet includes an exceptionally rare magnetic null line $\lambda$ 9743.61~\AA\ that has zero Land\'e factor and can be used to constrain Ti abundance and non-magnetic broadening.

Our analysis of the Y-band Ti~{\sc i} lines in the spectrum of \au\ followed the methodology described by \citet{shulyak:2014,shulyak:2017} and \citet{kochukhov:2017c}. We calculated theoretical intensity spectra with the help of the polarized radiative transfer code SYNMAST \citep{kochukhov:2010a} assuming a uniform radial magnetic field. The line list for these calculations was obtained from the VALD data base \citep{ryabchikova:2015} with the van der Waals damping parameters calculated according to \citet{barklem:2000b}. A solar metallicity MARCS \citep{gustafsson:2008} model atmospheres with $T_{\rm eff}=3700$~K \citep{houdebine:2016,afram:2019,plavchan:2020} and $\log g=4.5$ cm\,s$^{-2}$ was used. This surface gravity is close to the $\log g$ range 4.39--4.48 implied by different estimates of the stellar mass and radius given by \citet{plavchan:2020}. The microturbulent velocities are known to be relatively weak in M-dwarf atmospheres. Here we adopted $\xi_{\rm t}=0.25$~\kms\ according to the calibration provided by the three-dimensional hydrodynamic simulations by \citet{wende:2009}.

Several previous Zeeman broadening analyses of M dwarfs \citep{johns-krull:1996,johns-krull:2000,shulyak:2017,shulyak:2019} have demonstrated the need of multi-component field strength distributions for achieving satisfactory line profile fits. Here we approximated this distribution by a combination of $B=0$, 2, and 4~kG spectral components, added with the filling factors $f_0$, $f_2$, and $f_4$ ($f_0+f_2+f_4=1$). The filling factor $f_i$ represents the fraction of the stellar surface occupied by the field $B_i$. The same model atmosphere was used for calculating spectral contributions of all three components. Similar field strength distribution parameterizations are commonly used by the Zeeman broadening studies of M dwarfs and T Tauri stars \citep{johns-krull:1999,johns-krull:2000,johns-krull:2007,yang:2008,yang:2011,lavail:2019}. In our case, the free parameters fitted to the observed Ti~{\sc i} line profiles include the Ti abundance, the projected rotational velocity $v_{\rm e}\sin i$, the radial velocity $V_{\rm r}$, and the two filling factors $f_2$ and $f_4$. The individual continuum scaling factors were also allowed to vary in order to compensate small continuum normalisation errors \citep[e.g.][]{shulyak:2017}. The field-free filling factor $f_0=1-f_2-f_4$ and the mean field modulus \bs\,$\equiv$\,$\sum B_i f_i$\,$=$\,$2\cdot f_2+4\cdot f_4$ are dependent parameters derived from $f_2$ and $f_4$.

The parameter optimization was carried out with the help of the Markov chain Monte Carlo (MCMC) sampling algorithm implemented in the SoBAT set of routines \citep{anfinogentov:2020} written in IDL. The advantage of using a MCMC method compared to the traditional least-squares fitting is the possibility of incorporating prior constraints (e.g. the requirements $0\le f_i\le1$ and $\sum f_i=1$) through a rigorous Bayesian approach and obtaining robust error bars that account for all relevant parameter correlations. In this analysis of \au\ we ran the MCMC walkers for $10^5$ steps after the initial $10^4$ step burn-in stage. Uniform priors were assigned on each parameter. The best-fit parameter values were calculated as the median of the posterior distributions. The error bars were obtained from the same distributions, using the 68\% confidence interval.

\begin{figure*}[!th]
\fifps{\hsize}{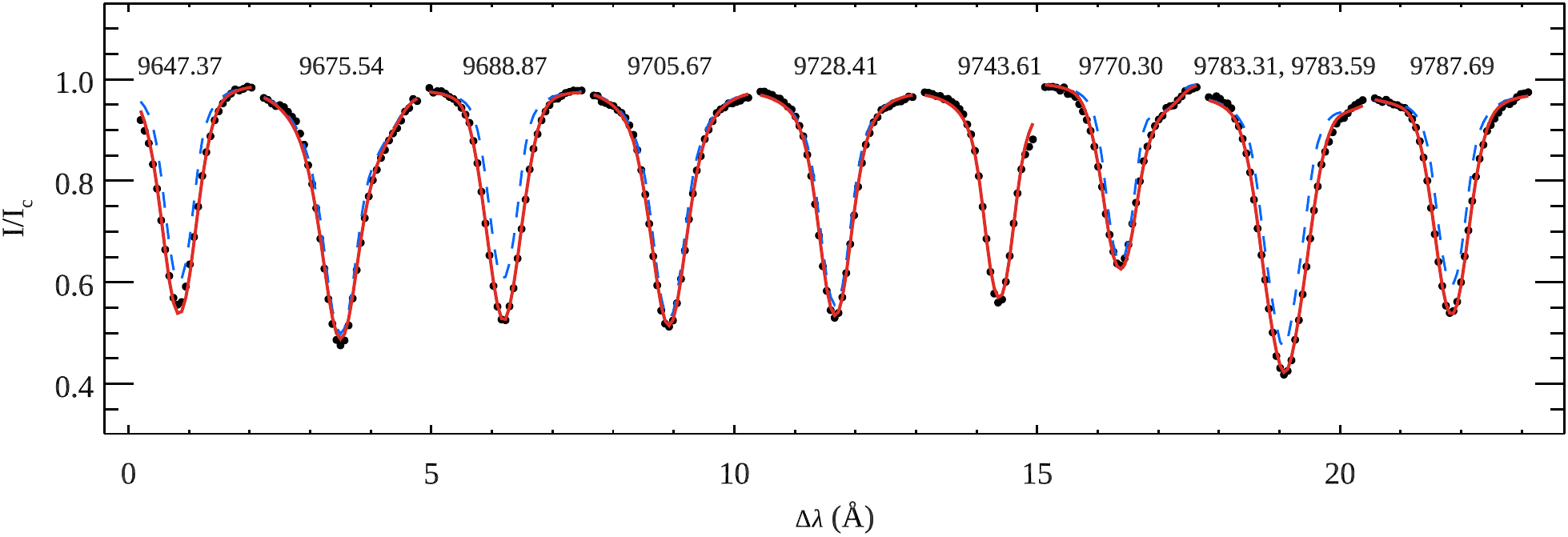}
\caption{Comparison of the observed (symbols) and computed (lines) profiles of nine near-infrared Ti~{\sc i} lines in the Y-band spectrum of \au. The solid line shows the best-fit magnetic model spectrum. The dashed line illustrates theoretical profiles calculated without magnetic field. The central wavelength of each transition, in \AA\ units, is indicated above the corresponding line profile.}
\label{fig:ti}
\end{figure*}

\begin{figure}[!th]
\fifps{\hsize}{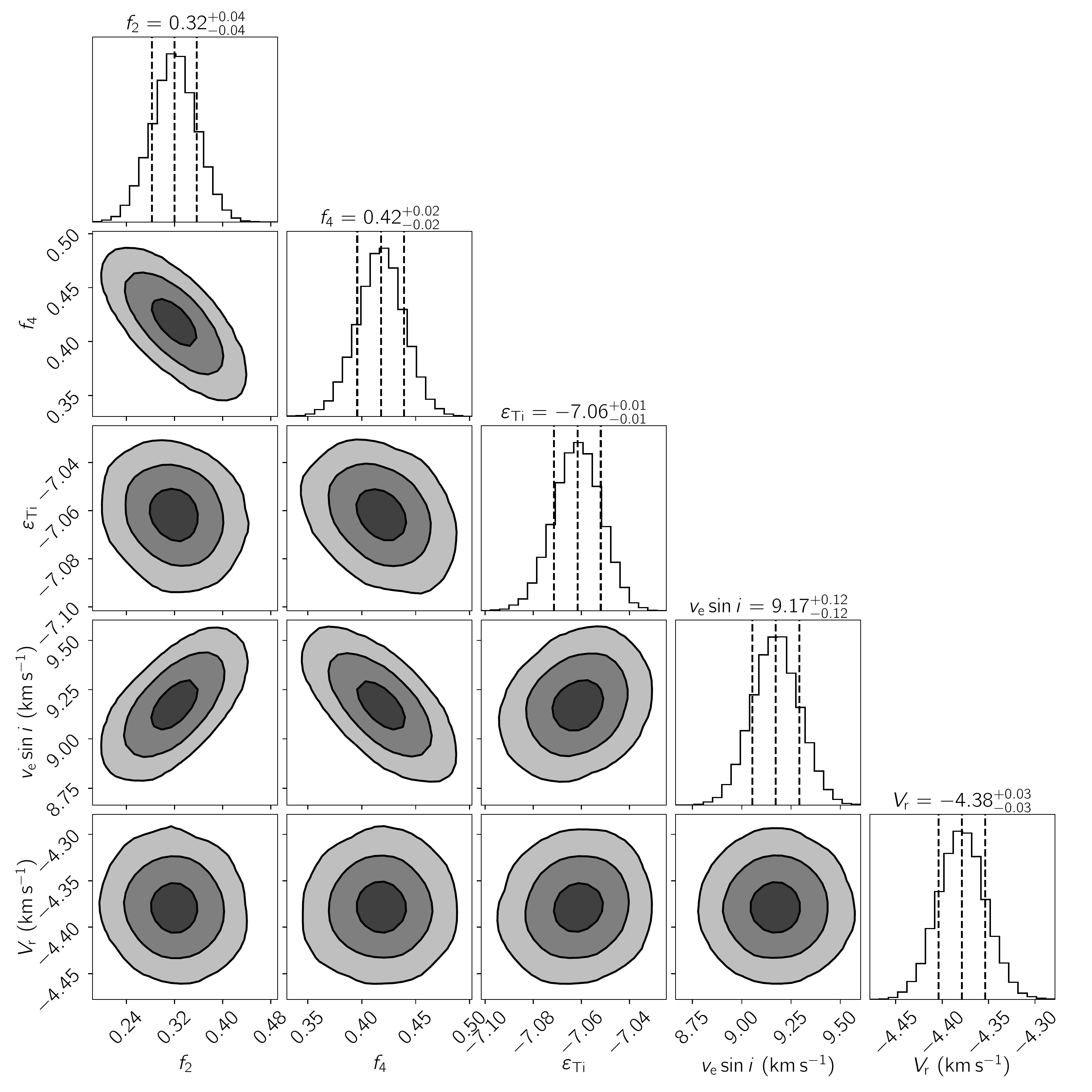}
\caption{Marginalized posterior distributions of independent parameters fitted to the Y-band Ti~{\sc i} lines. Contours correspond to 1, 2, and 3-$\sigma$ confidence levels.}
\label{fig:mcmc1}
\end{figure}

\begin{figure}[!th]
\fifps{0.8\hsize}{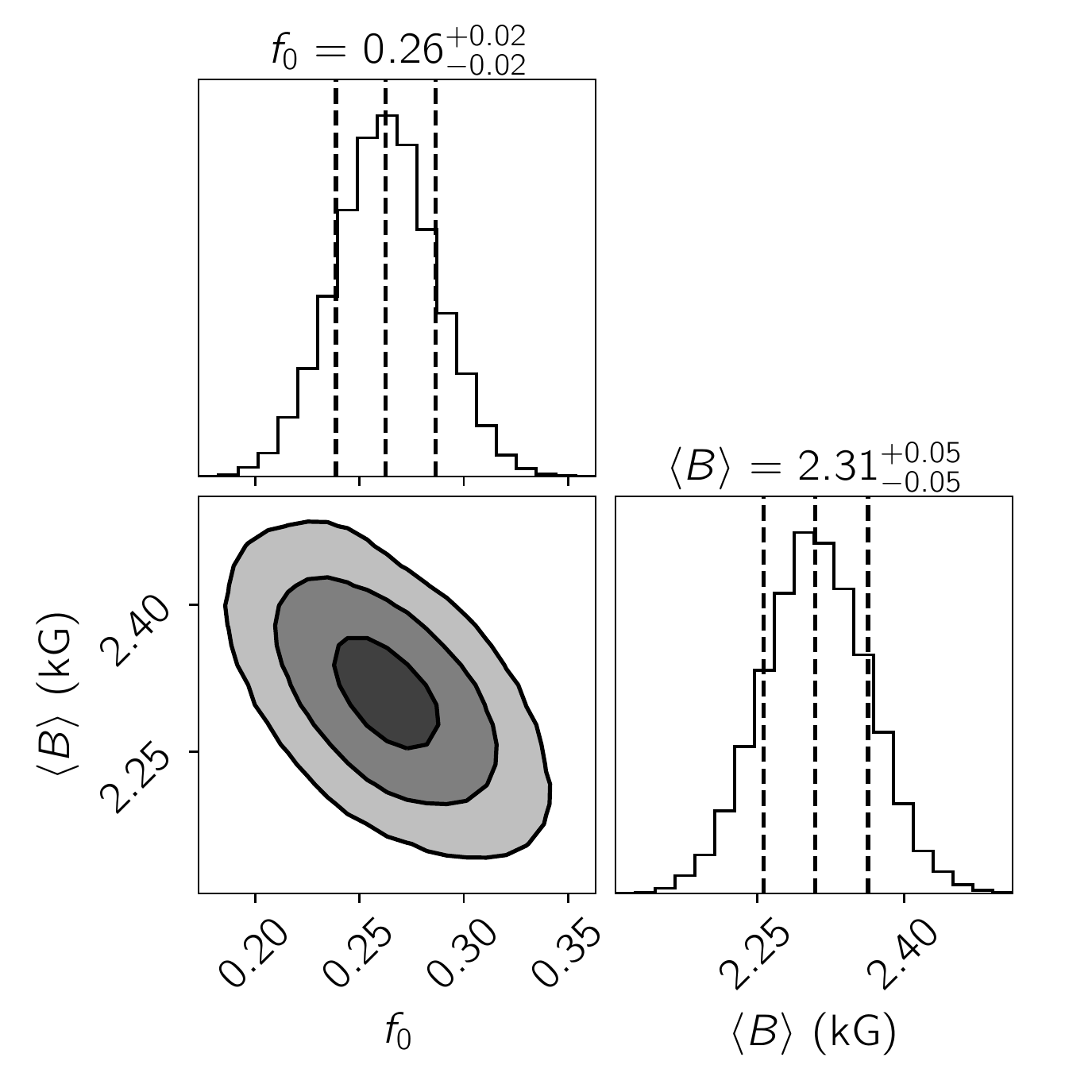}
\caption{Marginalized posterior distributions of the filling factor of field-free regions $f_0$ and the mean field modulus \bs\ derived from the Y-band Ti~{\sc i} lines.}
\label{fig:mcmc2}
\end{figure}

Our final theoretical fit to the Ti~{\sc i} lines in the Y-band spectrum of \au\ is shown in Fig.~\ref{fig:ti}. There is a clear evidence of both Zeeman broadening and magnetic intensification, with the Ti~{\sc i} $\lambda$ 9688.87~\AA\ line showing the strongest effect. The marginalized posterior distributions of independently fitted parameters are shown in Fig.~\ref{fig:mcmc1}. The corresponding distributions are illustrated in Fig.~\ref{fig:mcmc2} for the dependent parameters $f_0$ and \bs. Both of these plots were produced with the help of the Corner Python package \citep{foreman-mackey:2016}. The marginalized posterior distributions reveal several significant parameter correlations, for example between $f_2$ and $f_4$ and between both filling factors and $v_{\rm e}\sin i$ (Fig.~\ref{fig:mcmc1}). These correlations would normally be ignored by classical least-squares fits, leading to underestimated uncertainties. The correlation of filling factors suggests that using a finer grid of independently fitted magnetic components is not justified by the data. We have also verified that adding $B=6$~kG component does not result in a perceptible improvement of the fit quality and is not supported by the formal criteria, such as the Bayesian Information Criterion \citep{sharma:2017}.

\begin{table}[!t]
\centering
\caption{Magnetic field characteristics of \au.}
\label{tab:mag}
\begin{tabular}{lr}
\hline
\hline
Parameter &  Value~~~~ \\       
\hline
\multicolumn{2}{c}{\it Global magnetic field from LSD Stokes $V$} \\
$\langle B_V \rangle$ (kG) & 0.088 \\
$E_{\rm pol}$ (\%) & 91.2 \\
$E_{\ell=1}$ (\%) & 70.2 \\
$E_{|m|\ge\ell/2}$ (\%) & 96.0 \\
\hline
\multicolumn{2}{c}{\it Axisymmetric dipolar field from LSD Stokes $QU$} \\
$B_{\rm d}$ (kG) & 1.8--2.2 \\
\hline
\multicolumn{2}{c}{\it Mean field modulus from Y-band Ti~{\sc i} lines} \\    
\bs\ (kG) & $2.31\pm0.05(\pm0.2)$ \\
$\log (N_{\rm Ti}/N_{\rm tot})$ & $-7.06\pm0.01$ \\
$v_{\rm e}\sin i$ (\kms) & $9.17\pm0.12$ \\
\hline
\multicolumn{2}{c}{\it Mean field modulus from K-band Ti~{\sc i} lines} \\    
\bs\ (kG) & $2.08\pm0.02(\pm0.2)$ \\
$\log (N_{\rm Ti}/N_{\rm tot})$ & $-7.08\pm0.01$ \\
$v_{\rm e}\sin i$ (\kms) & $9.24\pm0.11$ \\
\hline
\end{tabular}
\tablecomments{The error bars given for \bs\ in brackets indicate systematic uncertainties.}
\end{table}

Results of our spectrum synthesis analysis of the Y-band Ti~{\sc i} lines are summarised in the middle part of Table~\ref{tab:mag}. Specifically, we found the mean field modulus \bs\,$=$\,$2.31\pm0.05$~kG. These relatively tight error bars correspond to the formal fitting errors obtained using a fixed set of model atmosphere parameters and a particular field strength parameterization. We estimated the systematic uncertainties by repeating the MCMC parameter inference changing $T_{\rm eff}$ by $\pm100$~K, $\log g$ by $\pm0.1$~dex, and $\xi_{\rm t}$ by $\pm0.25$~\kms. We also studied the effect of choosing a different field strength parameterization by using a three-component model with $B=1$, 3, and 5~kG \citep[e.g.][]{johns-krull:1999}. The outcome of this series of tests showed that \bs\ can increase by up to 0.2~kG and can be reduced by 0.1~kG whereas other parameters remain consistent within the error bars with the values reported in Table~\ref{tab:mag}.

The projected rotational velocity derived here is somewhat smaller than $v_{\rm e}\sin i$\,=\,9.3--9.7~\kms\ estimated by \citet{torres:2006} and \citet{houdebine:2010}. This difference is to be expected since we account Zeeman broadening in our spectroscopic modeling. On the other hand, one can obtain $v_{\rm e}=7.8\pm0.3$~\kms\ from $P_{\rm rot}$ and interferometric measurement of the stellar radius $R=0.75\pm0.03 R_\odot$ \citep{white:2019}. This is smaller than any spectroscopic values of $v_{\rm e}\sin i$ reported for \au\ in the literature. This discrepancy might be explained by a contribution of another broadening mechanism, for example enhanced macroturbulence, to the line profiles of this star. We have verified that our best-fitting $v_{\rm e}\sin i$ can be reduced down to 8.1~\kms\ if the radial-tangential macroturbulent broadening $\zeta_{\rm t}\approx4.5$~\kms\ is assumed for the analysis of Ti~{\sc i} lines. Nevertheless, this has no impact on the derived mean magnetic field strength or the filling factors of individual magnetic components.

\subsubsection{K-band analysis}
\label{sec:kband}

We carried out spectrum synthesis analysis of the CRIRES K-band observation of \au\ with the same procedure as was applied above. Considering temporal variability associated with rotational modulation and intrinsic changes of the surface magnetic field that might have occurred during six years separating ESPaDOnS and CRIRES observations, and possible formation depth difference of the spectral diagnostics, we analyze atomic lines in the Y and K band independently. Due to a limited wavelength coverage of the K-band spectrum, only four neutral titanium spectral lines, $\lambda$ 22211.22, 22232.84, 22274.01, 22310.61~\AA, are available for modeling. All these spectral features have large effective Land\'e factors (1.6--2.1 for the first three lines and 2.5 for Ti~{\sc i}  22310.61~\AA) and are significantly affected by a kG-strength magnetic field. Owing to a lack of suitable magnetically insensitive lines in the observed wavelength region, we fitted the K-band spectrum with a Gaussian prior on $v_{\rm e}\sin i$ according to the results of Sect.~\ref{sec:yband}.

\begin{figure}[!th]
\fifps{\hsize}{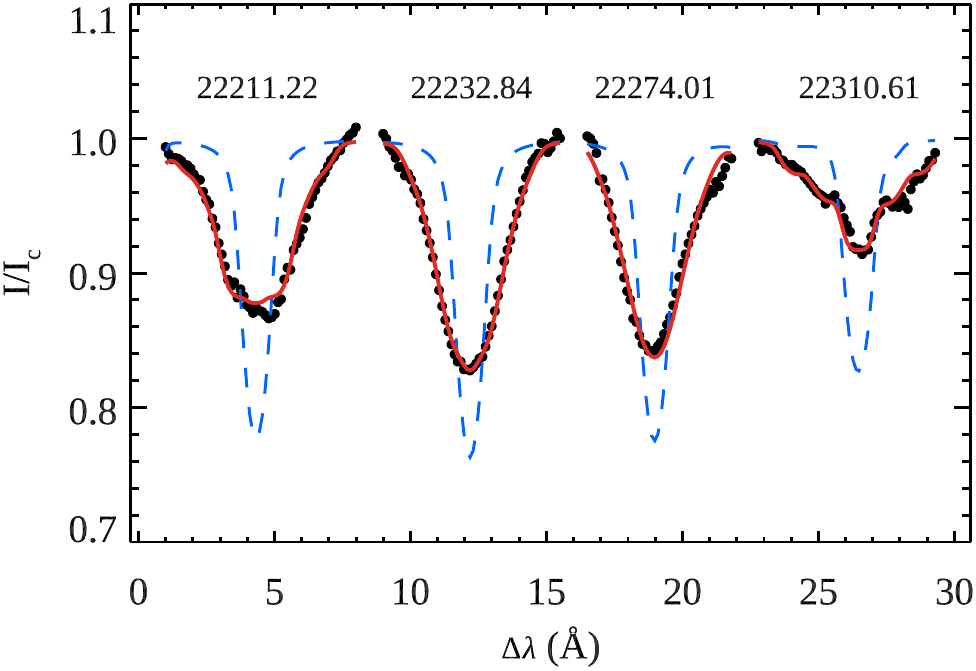}
\caption{Same as Fig.~\ref{fig:ti}, but for the K-band Ti~{\sc i} lines.}
\label{fig:ti1}
\end{figure}

\begin{figure}[!th]
\fifps{0.8\hsize}{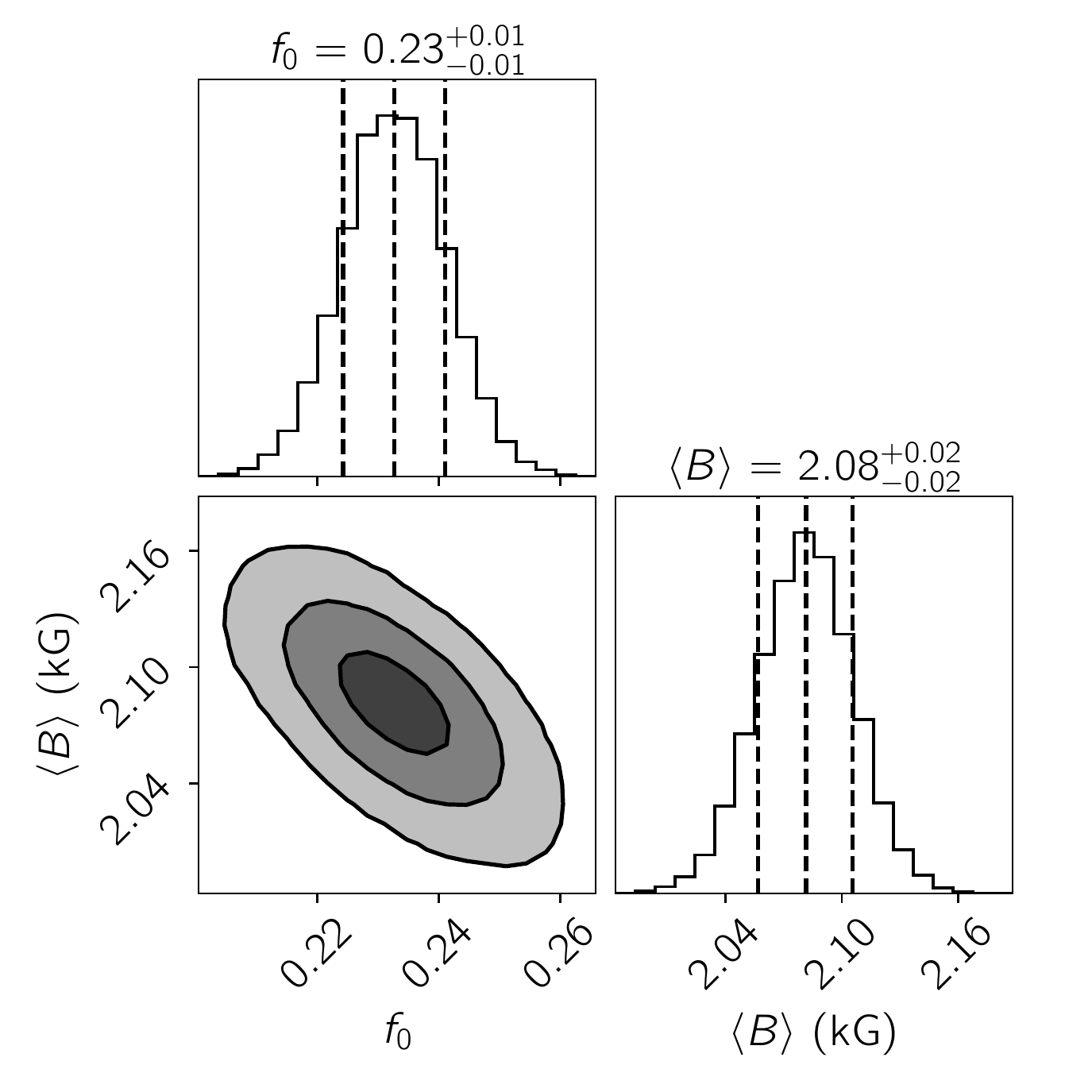}
\caption{Same as Fig.~\ref{fig:mcmc2}, but for the analysis based on the K-band Ti~{\sc i} lines.}
\label{fig:mcmc3}
\end{figure}

Figure~\ref{fig:ti1} shows comparison of the observed Ti line profiles with the best-fit theoretical model. The magnetic broadening and profile distortion of the studied spectral lines are clearly evident, particularly for the Ti~{\sc i}  22310.61~\AA\ line. The marginalized posterior distributions of $f_0$ and \bs\ obtained with the MCMC sampling are shown in Fig.~\ref{fig:mcmc3}. The K-band analysis results are reported in the bottom part of Table~\ref{tab:mag}. 

The CRIRES spectrum yields \bs\,$=$\,$2.08\pm0.02$~kG. This estimate of the mean field modulus is not affected by possible errors of $T_{\rm eff}$, $\log g$, and $\xi_{\rm t}$. At the same time, \bs\ increases to $2.32\pm0.04$~kG if the fitted field strength distribution is changed from $B$\,$=$\,0, 2, 4~kG to $B$\,$=$\,1, 3, 5~kG. This indicates a systematic error of $\approx$\,0.2~kG for \bs. The filling factor of non-magnetic regions, $f_0=0.23\pm0.01$, derived from the K-band Ti~{\sc i} lines in the framework of the former field strength distribution model, agrees with the Y-band results ($f_0=0.26\pm0.02$). Likewise, we found no improvement of the fit quality when the field strength distribution was extended to include the $B=6$~kG component.

\section{Discussion}
\label{sec:disc}

In this paper we presented a comprehensive analysis of the surface magnetic field of the young M-dwarf star \au. We studied both the polarization diagnostics linked to the large-scale stellar magnetic field and the Zeeman broadening and intensification of intensity line profiles, which captures the total magnetic flux. Using high-resolution Stokes parameter observations collected with two different spectropolarimeters, we systematically detected circular polarization in the photospheric line profiles with the help of the LSD multi-line technique. We reported 12 mean longitudinal magnetic field measurements spanning the range from $-34$ to $69$~G. These measurements correspond to the spectra acquired on nine individual nights during two observing epochs (2005--2006 and 2010) separated by about four years. No qualitative change of the characteristics of the global field of \au, such as changes of the Stokes $V$ profile morphology or magnitude of \bz, was found between these two epochs. The only other longitudinal field measurement available for \au\ was recently obtained by \citet{martioli:2020} based on near-infrared spectropolarimetric observations carried out in June 2019. These authors derived \bz\,$=$\,46~G, which is in agreement with the \bz\ range of our optical measurements. The circular polarization profiles of \au\ exhibit a strong rotational modulation, alternating in shape from symmetric to anti-symmetric. This shape variation, and the mere presence of symmetric crossover Stokes $V$ profiles, suggests a non-axisymmetric global magnetic field topology.

We also detected a linear polarization signal in the photospheric line profiles of \au. This signal is manifested as a nearly stationary signature in Stokes $U$. The shape of this signature and its dependence on the spectral line parameters indicates that it arises due to the Zeeman effect. Weak linear polarization was previously detected in the active M dwarf AD~Leo \citep{lavail:2018} and a few other more massive active late-type stars \citep{kochukhov:2011,rosen:2013}. These studies reported the ratio of the Stokes $V$ to Stokes $QU$ amplitude to be about 5--13. Here we found a much smaller ratio of 2--3, making linear polarization of \au\ unusually strong for the Stokes $V$ signal observed in this star. This may be due to the equator-on orientation of the star, which facilitates cancellation of the circular polarization signatures produced by any global field component anti-symmetric with respect to the stellar equator.

Preliminary ZDI analysis based on the subset of Stokes $V$ profiles collected in 2010 reveals a weak non-axisymmetric magnetic field configuration with a peak field strength of 184~G and a mean field strength of $\langle B_V \rangle$\,$=$\,88~G. These global field characteristics of \au\ are similar to some of the fields reconstructed with ZDI for early-M dwarfs \citep{donati:2008,kochukhov:2019a}, although a combination of the dominant poloidal field with a large degree of non-axisymmetry of this field component is relatively uncommon in these stars.

Consideration of linear polarization completely changes our perspective on the global magnetic field of \au. We showed that the large linear-to-circular polarization amplitude ratio observed for this star can be reproduced by adding $\approx$\,2~kG axisymmetric dipolar field to the ZDI map reconstructed from Stokes $V$ profiles. This field component is not observed in circular polarization due to the cancellation effect but gives rise to an observable linear polarization. The presence of a strong axisymmetric dipolar field makes magnetic field configuration of \au\ more similar to the large-scale fields found in mid- and late-M dwarfs \citep{morin:2008,morin:2010} or to the fields of T~Tauri stars with small radiative cores \citep{donati:2008b,donati:2010}. The latter similarity is probably not too surprising given the youth of \au\ and its pre-main-sequence evolutionary status \citep{plavchan:2020}.

\mybf{The problem of underestimation of the strength of axisymmetric magnetic field component is not specific to our ZDI analysis of \au\ but, in principle, affects any Stokes $V$ investigation of an equator-on star. For this cancellation effect to be significant, the stellar inclination angle must be  close to 90\degr. This situation is rare for a random star but common for transiting exoplanet hosts \citep{fares:2010} and detached eclipsing binaries \citep{kochukhov:2019a}. Thus, interpretation of ZDI results obtained for these objects should take into account possible presence of a strong axisymmetric field not captured by the circular polarimetric diagnostics.}

Our analysis of the global field topology of \au\ is based on a data set with a rather sparse phase coverage and therefore should be considered preliminary. It can be significantly improved with the help of new full Stokes vector observations with a dense phase coverage. We emphasize that linear polarization spectra provide a crucial piece of information on the large-scale field of this star that should not be neglected. Future ZDI investigations will have to incorporate the Stokes $Q$ and $U$ data in the magnetic inversions in order to provide a complete characterization of the global field of \au.

Polarimetric investigation of the global magnetic field of \au\ was complemented by the measurement of the mean field modulus from intensity spectra. This analysis provides information on the total magnetic field of the star, including the small-scale magnetic component that cannot be assessed with polarimetric methods. We applied detailed magnetic spectrum synthesis modeling to the two groups of near-infrared Ti~{\sc i} lines located in the Y and K bands. From the first set of lines we inferred \bs\,$=$\,$2.3\pm0.2$~kG whereas the second one yielded \bs\,$=$\,$2.1\pm0.2$~kG. In both cases, the best description of the observed line profiles was achieved by combining at least three spectral contributions corresponding to magnetic fields ranging from 0 to 4~kG in strength. In the context of this modeling approach, we inferred that about 23--26\% of the stellar surface is field-free.

Two previous studies analysed Zeeman broadening in \au\ spectra with a two-component approach, which assumes that a fraction $f$ of the stellar surface is covered by the field $B$ and the rest is non-magnetic. \citet{saar:1994a} measured \bs\,$=$\,$Bf=2.3$~kG from K-band Fourier transform spectra, in excellent agreement with our results. On the other hand, \citet{afram:2019} determined \bs\,$=$\,3.2--3.4~kG from different sets of atomic and molecular lines in the optical. \mybf{Their selection of atomic lines included three Ti~{\sc i} transitions at $\lambda$ 8364--8385~\AA\ with parameters similar to the Y-band titanium lines studied in our paper.} The authors estimated the error of their measurement to be about 0.6~kG, making these results incompatible with the outcome of our study. The analysis by \citet{afram:2019} relied on one of the ESPaDOnS spectra (July 16, 2005) studied in Sect.~\ref{sec:lsd}. Although we used a different ESPaDOnS observation from 2006 for the Y-band \bs\ determination, direct comparison of these two spectra reveals no difference of the strengths and/or shapes of the Ti~{\sc i} lines at $\lambda$ 9640--9790~\AA\ employed in our work. Thus, the discrepant field strength measurement reported by \citet{afram:2019} cannot be attributed to intrinsic variation of the star.

The mean field modulus derived by us for \au\ coincides with \bs\,$=$\,2.2~kG suggested by \citet{cranmer:2013} based on the field equipartition arguments. This lends support to the model of turbulent coronal heating developed by these authors to explain X-ray, radio, and millimeter observations of \au.

Comparison of the mean global field strength estimated from the tentative Stokes $V$ ZDI with the mean field modulus inferred from Zeeman broadening suggests that most of the magnetic field energy of \au\ is concentrated on small scales. The ratio $\langle B_{V} \rangle/\langle B \rangle \approx 4$\% is typical of early-M dwarfs with weak, complex large-scale fields \citep{kochukhov:2019a}. However, the 2~kG axisymmetric dipolar field favoured by the linear polarization data implies the disk-averaged global field of $\langle B_{QU} \rangle \approx1$~kG and $\langle B_{QU} \rangle/\langle B \rangle \approx 46$\%. This ratio of the global-to-total field strength is unusually high for an M dwarf but has been observed for some T~Tauri stars \citep{lavail:2019}.

The information on the magnetic field of \au\ provided in this study will be essential for analyses of star-planet magnetic interaction and for modeling stellar forcing on the planetary atmosphere. Given the planet's orbital distance of $\approx$\,19 stellar radii, it is likely to be located beyond the Alfv\'en surface. This makes detailed MHD stellar wind simulations \citep[e.g.][]{vidotto:2013,cohen:2014} necessary for estimating the pressure exerted by the stellar wind and extended magnetosphere on the planet. Such simulations are enabled by the constraints on global magnetic field reported here. On the other hand, the accurate measurement of the total magnetic flux constrains coronal heating and wind acceleration due to Alvf\'en wave turbulence \citep{garraffo:2016}.

\acknowledgements{
We thank Lisa Ros\'en for her contribution to the reduction and preliminary analysis of the HARPSpol observations of \au\ and Mathias Zechmeister for reducing the CRIRES spectrum of this star. We also thank Paul Barklem for calculating the van der Waals damping parameters for the spectral lines studied in this paper.

This study was supported by the Swedish Research Council and the Swedish National Space Agency. Based on observations collected at the European Southern Observatory (programs 85.D-0296 and 89.C-0173). Also based on observations obtained at the CFHT which is operated by the National Research Council of Canada, the Institut National des Sciences de l'Univers (INSU) of the Centre National de la Recherche Scientifique (CNRS) of France, and the University of Hawaii. 

This work has made use of the VALD database, operated at Uppsala University, the Institute of Astronomy RAS in Moscow, and the University of Vienna.
}

\newpage

\vspace{5mm}
\facilities{ESO:3.6m (HARPSpol), ESO:VLT (CRIRES), CFHT (ESPaDOnS)}

\mybf{
\software{REDUCE \citep{piskunov:2002}, iLSD and SYNMAST \citep{kochukhov:2010a}, InversLSD \citep{kochukhov:2014}, SoBAT \citep{anfinogentov:2020}, Corner \citep{foreman-mackey:2016}}
}

\end{document}